\documentclass[%
 reprint,
 amsmath,amssymb,
 aps,
pra,
floatfix,
]{revtex4-2}

\usepackage[utf8]{inputenc}

\usepackage{graphicx}
\usepackage{dcolumn}
\usepackage{bm}
\usepackage{soul}

\usepackage[dvipsnames]{xcolor}
\input{macros.sty}
\usepackage{comment}
\usepackage{stfloats}
\begin{document}

\title{Phase diagram and spectral
function of the two-dimensional disordered Bose-Hubbard model: A real-space dynamical mean-field theory analysis}

\author{Bastian Schindler}
\email{bastian.schindler@uni-hamburg.de}
\author{Renan da Silva Souza}%
\email{souza@itp.uni-frankfurt.de}
\author{Walter Hofstetter}
\email{hofstett@physik.uni-frankfurt.de}
\affiliation{%
Goethe-Universität, Institut für Theoretische Physik, 60438 Frankfurt am
Main, Germany
}%

\date{\today}

\begin{abstract}
We numerically investigate the two-dimensional Bose-Hubbard model with local onsite disorder, where the competition between disorder and short-range interactions leads to the emergence of a Bose glass (BG) phase between the Mott insulator (MI) and superfluid (SF) phases.~In order to analyze the inhomogeneous system we employ real-space bosonic dynamical mean-field theory (RBDMFT) and perform an ensemble average over disorder realizations.~To distinguish the MI from the BG phase, we compare the Edwards-Anderson order parameter and the compressibility with the energy-gap condition.~To identify the insulator to SF transition, we apply a percolation analysis to the condensate order parameter.~In qualitative accordance with the theorem of inclusions we always find an intermediate BG phase between the SF and MI.~
However, the quantitative comparison indicates significant deviations between the MI to BG phase boundary expected in the thermodynamic limit and the one obtained for a finite system size.~Additionally, RBMDFT is capable of reliably calculating spectral information throughout the phase diagram.~Analyzing the spectral function reveals evidence for analytically predicted damped localized modes in the dispersion relation in the strong-coupling regime.
\end{abstract}

\maketitle


\section{\label{sec:intro}Introduction}
Investigating the competition between disorder and contact interactions is essential for understanding insulating and glassy phases in strongly correlated quantum systems.~This question has been extensively addressed both experimentally and theoretically using systems of ultracold quantum gases in optical lattices with disorder \cite{sanchez2010disordered,shapiro2012cold}.~A particular example where this interplay is realized is the Bose glass (BG) phase which emerges when interacting bosonic quantum particles are subjected to a random external potential.~Originally predicted by Fisher \textit{et al.}~\cite{fisher_boson_1989}, the Bose glass phase is an insulating phase distinguished from the Mott insulator (MI) by its gapless excitation spectrum and finite compressibility and from the superfluid (SF) phase by the absence of long-range coherence.~Despite considerable advances, realizing and detecting the Bose glass phase remains a challenge.~Early experiments have observed the transition from a Mott insulating state to a state with a flat density of excitations and no phase coherence, which suggested the formation of Bose glass phase \cite{fallani2007ultracold}.~Subsequent work probed the Bose glass to superfluid transition by tracking vortex excitations generated during a disorder quench, revealing the existence of superfluid puddles in the Bose glass phase \cite{meldgin2016probing}.
More recently, time-of-flight imaging in two-dimensional quasicrystals of ultracold bosons has been used to study the same transition \cite{yu_observing_2024}.~However, the latter approach restricts the identification of the Bose glass phase to interaction strengths where no Mott insulating state can occur.~
Motivated by these limitations, here we theoretically investigate the characterization of the Bose glass phase using different observables and study its transitions to both the superfluid and Mott insulating states.

The Bose glass phase is predicted to intervene between the Mott insulator and superfluid phases in the presence of disorder. The absence of a direct superfluid to Mott insulator transition in this case is rigorously derived in the thermodynamic limit using the theorem of inclusions \cite{pollet_absence_2009}, which states that no gapped to gapless transition is possible except for transitions of Griffith's type.
This implies that in the thermodynamic limit, the transition from the Mott to the Bose glass phase occurs when the energy gap $E_g$ for particle-hole excitations of the Mott phase in the clean case becomes smaller than the disorder bound $\Delta$ \cite{gurarie_phase_2009, soyler_phase_2011}.
Within these studies, this condition for the Bose glass phase was combined with Monte Carlo (MC) simulations to fully determine the two dimensional \cite{soyler_phase_2011} and three dimensional \cite{gurarie_phase_2009} phase diagrams of the disordered model, which are in agreement with the predictions of \cite{fisher_boson_1989}.
While this energy gap criterion is a useful benchmark for the expected behavior in the thermodynamic limit, it does not characterize a disordered model directly as it only relies on the energy gap of the clean system and the given disorder bound $\Delta$.
Its derivation exploits the fact that in an infinite system all possible - even exponentially rare - configurations of the disorder potential are realized.
The argument may therefore not be applicable for even very large finite systems. 
An alternative characterization is provided by the Edwards-Anderson order parameter (EAOP), in analogy to spin-glass theory.
Applied to the disordered Bose-Hubbard model the EAOP is defined in terms of the variance of the local occupations over different disorder realizations \cite{graham_order_2009, khellil_analytical_2016, khellil_hartreefock_2016}.
It was successfully applied in a numerical study using mean-field theory \cite{thomson_measuring_2016} and in recent experimental studies of the Bose glass phase \cite{abadal_probing_2020, koehn_quantum-gas_2025}.
It is closely related to the compressibility, which was recently measured experimentally \cite{russ2025compressibility}.
Additionally, both statistical \cite{bissbort_stochastic_2009, bissbort_stochastic_2010} as well as site-resolved \cite{buonsante_mean-field_2007, niederle_superfluid_2013} mean-field theories proved powerful tools to investigate the disordered Bose-Hubbard model in a non-perturbative way.
Another important aspect is the characterization of the Bose glass excitation spectrum which could also be used to identify this phase.
Using a strong-coupling expansion it was shown in \cite{souza_emergence_2023, souza_ultracold_2023} that the spectrum of the disordered Mott insulator features a broad distribution associated with damped-localized excitations. These excitations persist in the spectrum for strong disorder when the Mott phase is suppressed and a phase transition to the Bose glass phase takes place. Hence, it was argued in \cite{souza_emergence_2023, souza_ultracold_2023} that these correspond to single-particle excitations of the Bose glass. 

In the present study, we extend previous mean-field analysis by including the terms of second order in $1/z$, where $z$ is the lattice coordination number, which results in bosonic dynamical mean-field theory (BDMFT) \cite{byczuk_correlated_2008, hubener_magnetic_2009, hu_dynamical_2009, anders_dynamical_2011, proukakis_bosonic_2013}. 
Due to the inhomogeneous nature of the system, we use the site-resolved variant of the BDMFT equations named real-space BDMFT {(RBDMFT)}, which was, for example, successfully used to investigate magnetic order in the two-component Bose-Hubbard model \cite{li_tunable_2011}.
We introduce the model in the next section and outline the RBDMFT equations in the appendix. We consider a system of spinless ultracold bosons confined to a square optical lattice in the presence of an onsite disorder potential.
We account for the stochastic nature by an arithmetic average over disorder realizations, which is assumed to reproduce the ensemble average for sufficient sampling.
Using a modified version of the percolation analysis proposed in \cite{niederle_superfluid_2013}, we were able to successfully determine the superfluid to insulator phase boundary in agreement with previous Monte Carlo studies \cite{soyler_phase_2011}.
In conjunction with the Edwards-Anderson order parameter, we were able to fully construct the phase diagram for fixed disorder as well as for unit filling. Additionally, we show the evolution of the spectral function throughout the three different phases and discuss the good agreement between the spectral function obtained numerically using RBDMFT and the analytical results from afore-mentioned strong-coupling expansion~\cite{souza_emergence_2023}.

\begin{figure}[ht!]
\includegraphics[width=0.5\textwidth]{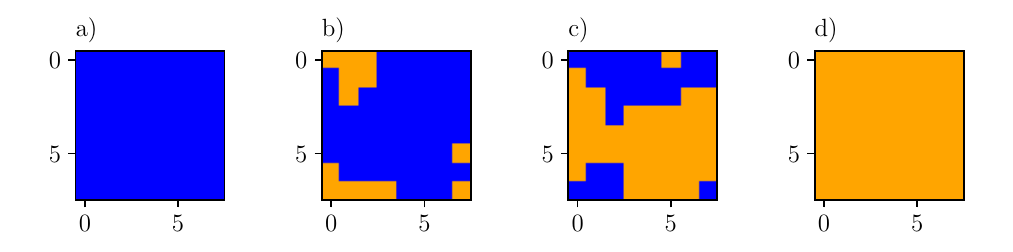}
\caption{\label{fig:discreteMap} Illustration of the percolation analysis for four different points in the low-temperature phase diagram of the disordered Bose-Hubbard model on an $8 \times 8$ square lattice with periodic boundary conditions and a fixed disorder strength of $\Delta / U = 0.5$ as well as hopping parameter $J / U = 0.03$. The discrete map $M_i$ is used with a cutoff $\phi_c = 0.05$ as defined in \eqref{eq:discreteMap}. Blue (orange) squares indicate a value $M_i = 0$ ($M_i = 1$).
Here (a) $\mu / U = 0.5$ and (b) $\mu / U = 0.65$ are classified as insulating realizations, whereas (c) $\mu / U = 0.65$, which is derived from a different disorder realization than (b), is categorized as a superfluid realization.  Lastly (d) $\mu / U = 0.9$ corresponds to a superfluid realization.
}
\end{figure}

\section{\label{sec:model}Disordered Bose-Hubbard Model}

We consider the Bose-Hubbard model with a diagonal onsite disorder term given by
\begin{equation}
    \label{eq:dBHM}
	\hat{H} = - J \sum_{\langle ij \rangle} \hat{b}^{\dagger}_i \hat{b}_j\ + \frac{U}{2} \sum_i \hat{n}_i (\hat{n}_i - 1) - \sum_i ( \mu -  \epsilon_i)\hat{n}_i ,
\end{equation}
which describes interacting spinless bosons in the lowest Bloch bands of an optical lattice \cite{jaksch_cold_2005, greiner_quantum_2002}.
Here $J$ is the (uniform) hopping amplitude, $U$ is the interaction strength originating from \textit{s}-wave scattering, and $\mu$ is the chemical potential, which is modulated by the random onsite potential $\epsilon_i \in [-\Delta / 2, \Delta /2]$ drawn from a uniform distribution bounded by the maximal disorder strength $\Delta$. 

Experimentally, disordered potential landscapes can be generated, e.g., using laser speckles \cite{clement2006experimental} or digital micromirror devices \cite{ren2015tailoring}.
The operators $\hat{b}_i$ and $\hat{b}^{\dagger}_i$ are the bosonic annihilation and creation operators, respectively, and $\hat{n}_i = \hat{b}^{\dagger}_i \hat{b}_i$ is the number operator on site $i$.

Within RBDMFT, the lattice Hamiltonian Eq.\,\eqref{eq:dBHM} is mapped to $N_s$ (number of sites) quantum impurity problems \cite{li_tunable_2011}, which are solved independently using an exact diagonalization of the resulting effective Anderson impurity Hamiltonian.
The self-consistency is closed using the lattice Dyson equation formulated in real space \cite{li_tunable_2011}, in contrast to BDMFT for translationally invariant systems, where the lattice Dyson equation is formulated in $\vec{k}$ space \cite{proukakis_bosonic_2013}.
This approach allows one to fully account for the position dependence of the disorder potential.
We introduce the relevant equations and explain the method in more detail in the appendix.

\section{\label{sec:results} Disordered Phase Diagram}

We first discuss the results for the $\mu / U$ vs $J / U$ phase diagram at fixed values of the strength of the disorder potential $\Delta / U = 0.5$ and $1.0$. Secondly, we compute the $\Delta / J$ vs $U / J$ phase diagram at unit filling in order to quantitatively compare our results to the MC study in \cite{soyler_phase_2011}. All calculations were performed for $\beta U=10^8$, where $\beta=1/T$ is the inverse temperature and we have assumed $k_{\rm B}=1$.
Thus, the results are expected to reflect the behavior of the system at zero temperature.

\begin{figure}[ht!]
\includegraphics[width=0.5\textwidth]{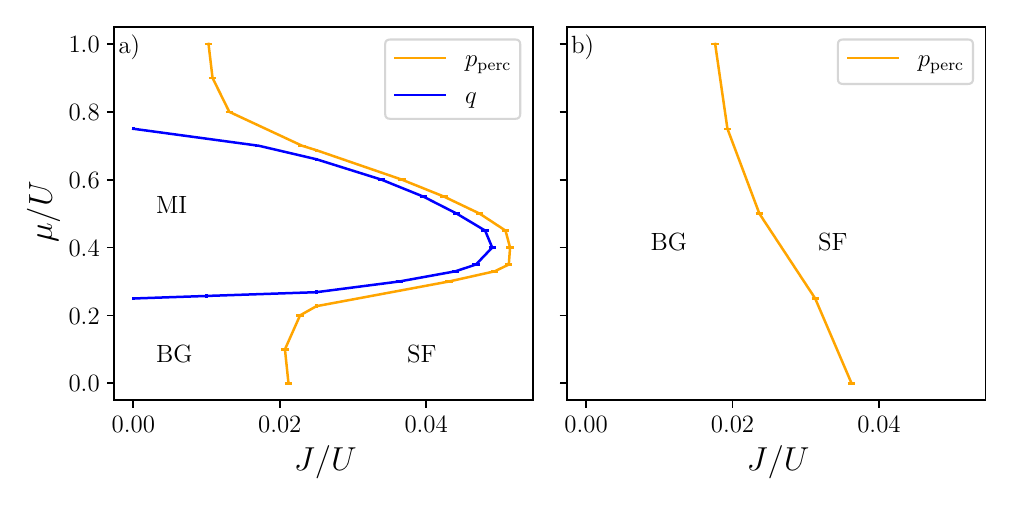}
\caption{\label{fig:fixedPD} 
$\mu / U$ vs $J / U$ phase diagram for a fixed disorder bound of $\Delta / U = 0.5$ in (a) and $\Delta / U = 1.0$ in (b) on an $8 \times 8$ lattice using $1536$ disorder realizations. 
The insulator to superfluid transition is obtained via the condition $\pperc = 0.5$ with $\phi_c = 0.05$.
A point in the phase diagram where the value of the EAOP is below $0.001$ is considered to be Mott insulating, which determines the MI to BG phase boundary.
For $\Delta / U  = 1.0$ no Mott insulating phase is present and the shape of the superfluid to Bose glass transitions changes significantly.}
\end{figure}

In order to characterize the superfluid to insulator transition, we employ a percolation analysis inspired by \cite{niederle_superfluid_2013}, where it was applied to the local occupations in site-resolved Gutzwiller calculations.
In our case we define the elements of the discrete map $M_i$ using the local condensate order parameter $\phi_i = \langle \hat{b}_i \rangle$:
\begin{eqnarray}
    M_i = 
	\begin{cases}
      1, & \text{for } \phi_i > \phi_c, \\
      0, & \text{else}. \\
    \end{cases}
    \label{eq:discreteMap}
\end{eqnarray}

Here, the critical value for local superfluidity $\phi_c$ is usually chosen to be approximately between $0.05$ and $0.1$.
This approach was already successfully applied within static mean field calculations of the disordered Bose-Hubbard model using a similar critical value~\cite{gupta2024strong}.
A selection of such maps for different parameter values of the Hamiltonian is shown in Fig.\ \ref{fig:discreteMap}.

For each disorder strength at a given point in the phase diagram, we count the number of disorder realizations for which the system contains at least one percolating cluster, defined as a connected set of SF sites ($M_i=1$) that spans between opposite sides of the lattice. Dividing this number by the total number of disorder realizations, we obtain the percolation probability $\pperc$.
If $\pperc > 50 \%$, we identify the investigated point in the phase diagram to be in a globally superfluid phase.

To distinguish the Bose glass phase from the Mott insulator, we mainly consider the EAOP of the disordered Bose-Hubbard model, given by the local variance of the occupation~\cite{graham_order_2009, khellil_analytical_2016, khellil_hartreefock_2016, thomson_measuring_2016}
\begin{eqnarray}
    q =  \frac{1}{N_r} \sum_r {(n_r - \bar{n})^2} =  \overline{n^2} - \bar{n}^2,
    \label{eq:eaop}
\end{eqnarray}
where $r = 1, \dots, N_r$ indexes the different realizations of disorder.
The overline denotes the ensemble average, which is approximated within the numeric implementation by the arithmetic average over the different realizations of disorder for each phase diagram point.
Practically, the EAOP according to Eq.\,\eqref{eq:eaop} will be site dependent for a finite system calculation, where $n$ is the occupation of a certain site of the disorder realization $r$.
As for an infinite number of realizations $N_r$ the EAOP is site independent, we average over all sites, which improves the statistics of the numerically determined EAOP.

The EAOP is closely related to the compressibility $\kappa = \partial_{\mu} \overline{n}$; in particular $\kappa = 0$ in the Mott insulator implies $q = 0$.
Within the numerical calculations, the compressibility $\kappa$ is calculated using a second RBDMFT run with a slightly shifted chemical potential $\mu + \delta \mu$, where $\delta \mu = 0.01$, for each disorder realization.
The result is then averaged over the disorder realizations, which is equivalent to the definition of $\kappa$.

Using the criteria outlined, we construct the $\mu / U$ vs $J / U$ phase diagram at a fixed  $\Delta / U = 0.5$ [see Fig.\ \ref{fig:fixedPD} (a)] in qualitative agreement with the predictions of \cite{fisher_boson_1989}.
In particular, we always find an intermediate Bose glass phase separating the Mott lobe from the superfluid phase. With increasing disorder strength the Mott lobe shrinks such that for $\Delta/U \gtrsim 1$ the Bose glass phase dominates the insulating region of the phase diagram as seen in Fig.\ \ref{fig:fixedPD} (b).

Having established that the EAOP allows for an accurate determination of the MI to BG phase transition, we now analyze the unit filling phase diagram where we vary both disorder and interaction strengths.
We compare the MI to BG phase boundary extracted from analyzing the EAOP, the compressibility, and the energy-gap criterion $E_g\leq \Delta$ defined in \cite{pollet_absence_2009}. 

\begin{figure}[h!]
\includegraphics[width=0.5\textwidth]{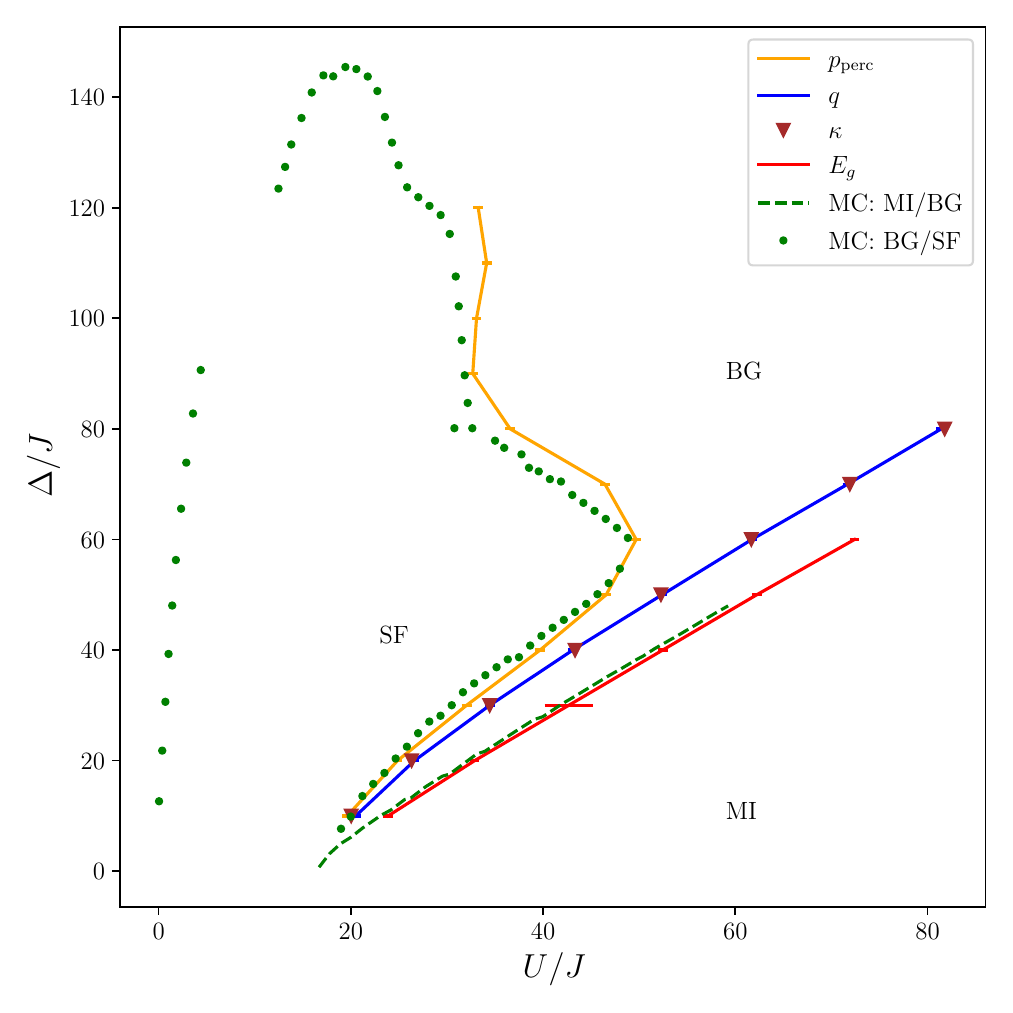}
\caption{\label{fig:unitfillingPD} 
$\Delta / J$ vs $U / J$ phase diagram for unit filling $\bar{n} = 1$ on an $8 \times 8$ lattice with an ensemble average using $1536$ disorder realizations.
The insulator to superfluid transition (orange line) is obtained via the condition $\pperc = 0.5$ with $\phi_c = 0.1$.
The MI to BG boundary has been determined using the EAOP (blue line) with a threshold of $0.002$ and compressibility (brown triangle) with a threshold of $0.01$, with both criteria giving the same result.
Conversely, the prediction from the energy-gap criterion (red line) using the spectral information of the homogeneous system within RBDMFT shows significant deviations.
Our results are compared to Monte Carlo (MC) simulations of the two-dimensional disordered Bose-Hubbard model at zero temperature published in \cite{soyler_phase_2011}, where the MI-BG transition (green dashes) was obtained using the energy-gap condition in the thermodynamic limit, and the SF-BG transition (green circles) was determined from the superfluid stiffness.}
\end{figure}
\begin{figure}[h]
\includegraphics[width=0.5\textwidth]{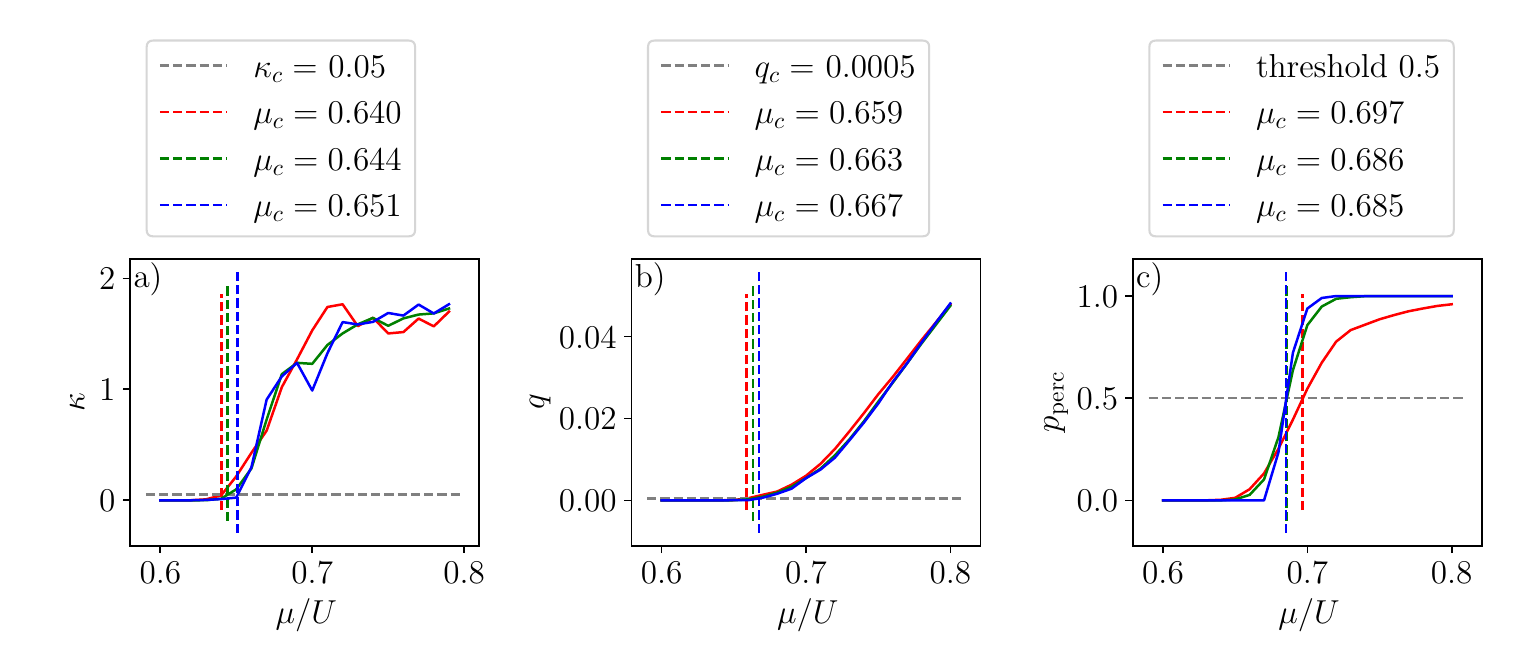}
\caption{\label{fig:finiteSize}
Plot of the compressibility $\kappa$ (a), EAOP $q$ (b) and percolation probability $p_{perc}$ (c) plotted for system sizes $4 \times 4$ (red), $8 \times 8$ (green) and $16 \times 16$ (blue) obtained for fixed $J/U = 0.025$.
}
\end{figure}

In Fig.\ \ref{fig:unitfillingPD} we see good quantitative agreement of our results with the MC results of \cite{soyler_phase_2011} for the superfluid to insulator transition.
Furthermore, the criterion for the BG to MI transition derived from the theorem of inclusions in \cite{gurarie_phase_2009, pollet_absence_2009} using the energy gap of the clean system $E_g \leq \Delta$ agrees perfectly with the MC calculations obtained in the thermodynamic limit.
When we turn towards $q$ and $\kappa$ calculated directly using RBDMFT for the finite-size disordered system, we see significant deviations in the location of the phase boundary (of the order of $10 \times U / J $) compared to the energy-gap criterion.
In fig. \ref{fig:finiteSize} we show the diagnostics used to determine the phase boundaries in Figs. \ref{fig:fixedPD} and fig. \ref{fig:unitfillingPD} for different system sizes ($4 \times 4$ to $16 \times 16$) in order to obtain insights about possible finite-size corrections.
Overall, within the accessible system sizes, we see only small deviations, which throughout the different phase-space points did not follow a consistent scaling. We thus conclude that finite-size effects are negligible and do not significantly impact the analysis of the phase diagram presented.
Nevertheless, while these deviations are apparently not explained by simple finite-size scaling, we still argue that the deviations originate from the fact that we consider a finite system size in our simulation. The existence of rare Lifshitz regions, in which many neighboring sites in a given disorder realization form a spatial domain with near-minimal potential, is only guaranteed in the thermodynamic limit. The presence of such rare regions is necessary in the justification of the theorem of inclusions and thus the energy-gap criterion. In finite systems, however, the probability for rare Lifshitz regions to occur is exponentially suppressed. Site-resolved Gutzwiller calculations \cite{niederle_superfluid_2013} support this picture, indicating that even for $100 \times 100$ lattices the Mott phase remains slightly larger than predicted by the energy-gap criterion.
Thus one will not observe a consistent finite-size scaling as long as these regions are effectively nonexistent.
An interesting point for further investigation is the question of whether there is a large but finite size for which the system (approximately) respects this energy-gap criterion.

\begin{figure}[h]
\includegraphics[width=0.5\textwidth]{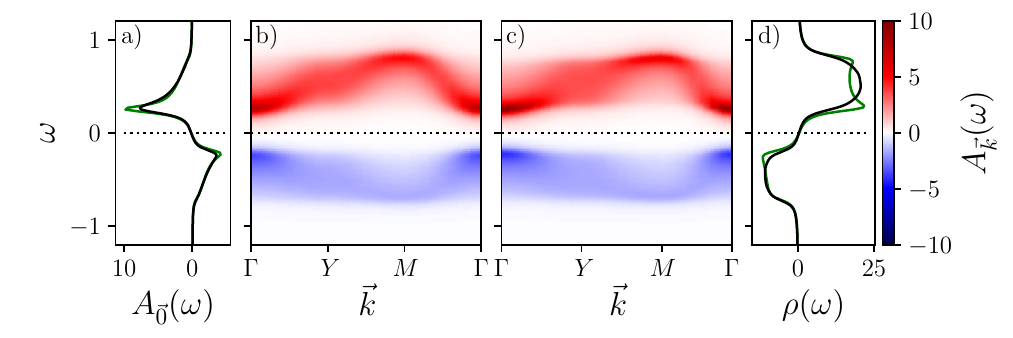}
\caption{\label{fig:spectralcomp}
Spectral function $A_{\kvec} (\omega)$ [Eq.\ \eqref{eq:spectral}] in the disordered Mott insulating phase. In all plots the chemical potential and the hopping amplitude are fixed at $\mu / U = 0.46$ and $J / U = 0.02$ and we used a numerical regularizer of $\izeroplus = 0.05$ as well as a disorder strength of $\Delta / U = 0.5$.
Here (a) depicts the $\kvec = \zerovec$ mode of the spectral function for both the numerical (black) and analytical (green) solution, (b) shows the dispersion relation as density plot of $A_{\kvec} (\omega)$ for the RBDMFT, and (c) shows that for the strong-coupling result. (d) Density of states $\rho(\omega) = \sum_{\kvec} A_{\kvec} (\omega)$ again for both the numerical and analytical solution.
The numerical solution is obtained using RBDMFT on a $16 \times 16$ square lattice, while the analytical solution is obtained by strong-coupling theory as in \cite{souza_emergence_2023, souza_ultracold_2023}.
}
\end{figure}

\begin{figure}
\includegraphics[width=0.5\textwidth]{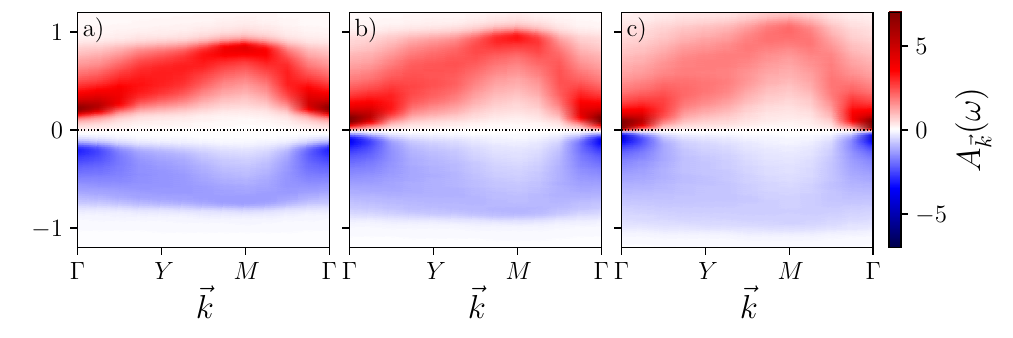}
\caption{\label{fig:spectral}
Evolution of the spectral function $A_{\kvec} (\omega)$ [Eq.\ \eqref{eq:spectral}] as a density plot at unit filling for $\Delta / J = 30$ from a disordered Mott insulator (a) $U / J = 50$ through the Bose glass phase in (b) $U /J = 37.58$ to the superfluid phase in (c) $U / J = 30.34$. 
}

\end{figure}

\section{\label{sec:results} Spectral Analysis}
A fundamental approach to characterizing the different phases of this system is the analysis of their excitation spectra. For the clean case, the spectrum has been extensively studied \cite{alon2005zoo, sengupta_mott-insulatorsuperfluid_2005,knap2010spectral, ejima2011dynamic, zaleski2012momentum, panas_numerical_2015, gremaud2016excitation}. Experimentally, low-lying excitations have been probed using techniques such as Bragg spectroscopy \cite{clement2009exploring, ernst2010probing, bissbort_detecting_2011} and lattice amplitude modulation \cite{stoferle2004transition,endres2012higgs}. The latter has enabled the observation of the continuous evolution of the clean MI spectrum from a gapped structure with stable quasiparticle and quasihole excitations to a spectrum featuring a gapless Goldstone mode and an amplitude (Higgs) mode in the clean SF phase \cite{endres2012higgs}. In contrast, the effect of disorder on the excitation spectrum has received considerably less attention.
Within our RBDMFT approach, the excitation spectrum becomes directly accessible through the exact-diagonalization impurity solver, without the need for numerical analytic continuation. To investigate it, we compute the site-resolved spectral function from the disorder-averaged lattice Green’s function in real-frequency space and Fourier transform it to $\kvec$ space,
\begin{equation}
    A_{\kvec} (\omega) = - \frac{1}{\pi} \imag \overline{G_{\kvec} (\omega + i 0^+)}.
    \label{eq:spectral}
\end{equation}
{Figure \ref{fig:spectralcomp} shows a comparison between the analytical strong-coupling results of \cite{souza_emergence_2023} and our numerical calculations, revealing strong qualitative and reasonably accurate quantitative agreement, given the restricted spectral resolution of the exact-diagonalization impurity solver. In particular, Figs.\ \ref{fig:spectralcomp}(a)-\ref{fig:spectralcomp}(d) provide clear evidence for the existence of damped-localized states in the excitation spectrum of the disordered Mott insulator. At $\vec{k}=\vec{0}$ [Fig.\ \ref{fig:spectralcomp}(a)], the spectral function displays two peaks with pronounced tails, separated by an energy gap, corresponding in the clean limit to the quasiparticle and quasihole excitations of the Mott phase. For finite disorder, across the high-symmetry points of the Brillouin zone, $\Gamma=(0,0)$, $Y=(0,\pi)$, and $M=(\pi,\pi)$ [Fig.\ \ref{fig:spectralcomp}(b)], the spectral function develops a broad background (blue and red shaded regions of approximate width $\Delta/U$), while remaining sharp only near the $\Gamma$ and $M$ points. The sharp features correspond to stable, long-lived excitations, whereas the broad regions correspond to signal damped, short-lived excitations without clear dispersion, leading to real-space localization. These damped-localized excitations arise from the presence of disorder. Due to the assumption of a vanishing variance of the onsite occupations ($q=0$), the results of \cite{souza_emergence_2023} are valid only within the MI phase and asymptotically close to the MI–BG transition from the MI side. In contrast, RBDMFT allows us to extend the spectral analysis across the entire phase diagram.}

To demonstrated this, in Fig.\,\ref{fig:spectral} we show the resulting spectral function for fixed disorder strength and different interaction values, plotted along the high-symmetry points of the Brillouin zone.~In the disordered MI phase [Fig.\,\ref{fig:spectral}(a)], the spectrum remains gapped but exhibits a broad background (red and blue shaded areas of approximate width $\Delta/U$), consistent with the behavior identified in Fig.\ \ref{fig:spectralcomp}.~Upon decreasing $U/J$, we find the BG spectrum [Fig.\,\ref{fig:spectral}(b)], which is gapless at the $\Gamma$ point, while retaining the broad, incoherent structure similar to the disordered MI spectrum.~For sufficiently small $U/J$, we obtain the disordered SF spectrum [Fig.\,\ref{fig:spectral}c)], where the spectral weight concentrates around $\Gamma$.~Overall, disorder leads to a significant loss of coherence in the single-particle spectrum and to the emergence of localized excitations.~These results can be experimentally verified using the approach of \cite{volchkov_measurement_2018}, where a Bose-Einstein condensate of $^{87}$Rb atoms in the hyperfine state $|1\rangle=|F=1,m_F=-1\rangle$ was loaded into a random potential generated by laser speckles acting only on atoms in the hyperfine states $|2\rangle=|F=2,m_F=1\rangle$. The spectral function was extracted from the transfer rate measured for the $|1\rangle\rightarrow|2\rangle$ transition driven by a radio-frequency field, according to Fermi's golden rule. Thus, our numerical spectral analysis is directly linked to recent experiments and, in particular, demonstrates the possibility of characterizing the Bose glass phase via spectral information.

For experimentally relevant higher temperatures ($T\gtrsim 0.1 U$), we expect that the incompressible MI is replaced by a normal insulator with thermally induced finite compressibility and that the transition to the BG phase becomes a crossover \cite{bissbort_stochastic_2010}. We expect that this crossover can be characterized via the spectral properties, specifically from the disorder-averaged zero-energy local density of states which vanishes in the normal insulator and is finite in the BG \cite{bissbort_stochastic_2010}. Moreover, recent quantum-gas-microscopy experiments \cite{koehn_quantum-gas_2025} show that the EAOP provides an unambiguous characterization of the BG phase even at $T \approx 0.1U$.

\section{\label{sec:conc}Conclusion}
We investigated the phase transitions of the two-dimensional disordered Bose-Hubbard model using RBDMFT at low temperatures and showed that in conjunction with a percolation analysis this approach is capable of accurately describing global superfluidity in disordered systems. 
We studied the MI to BG transition by comparing several diagnostics for the phase transition. We demonstrated that the energy-gap criterion, derived from the clean system
in the thermodynamic limit, significantly overestimates the extent of the BG phase in the disordered Bose-Hubbard model for a finite system size. 
Our analysis shows that the EAOP (or equivalently the compressibility) is the appropriate criterion for characterizing the MI to BG transition for finite systems.
This provides the foundation for further use of this quantity for identifying the BG phase, as has been done very recently experimentally using a quantum gas microscope \cite{koehn_quantum-gas_2025}.  
In addition, we have computed the spectral function of the disordered Bose-Hubbard model throughout the phase diagram, which shows good agreement with analytical predictions for the disordered Mott insulator.
This demonstrates a key advantage of the applied method as the RBDMFT approach enables a reliable characterization of spectral properties, providing supporting evidence for the existence of damped-localized excitations in the spectrum of the Mott insulating phase as predicted in \cite{souza_emergence_2023} and a transfer of spectral weight in the superfluid phase. This opens the possibilities for a variety of future analysis, which can be connected to recent experimental techniques.

\begin{acknowledgments}
The authors gratefully acknowledge the computing time
provided to them at the NHR Center NHR@SW at Goethe-
University Frankfurt. This is funded by the Federal Ministry
of Education and Research, and the state governments partici-
pating on the basis of the resolutions of the GWK for national high performance computing at universities \cite{nhr}.
The authors
gratefully acknowledge the Gauss Centre for Supercomputing
e.V. \cite{gcs} for funding this project by providing computing
time through the John von Neumann Institute for Computing
(NIC) on the GCS Supercomputer JUWELS \cite{juwels} at Jülich
Supercomputing Centre (JSC).
\end{acknowledgments}

\appendix
\section{\label{ap:RBDMFT}RBDMFT Equations}

DMFT was first successfully developed and applied to strongly correlated electron systems as a non-perturbative self-consistent theory \cite{metzner_correlated_1989}.
In order to accommodate inhomogeneous systems, a real-space extension of DMFT was implemented \cite{snoek_antiferromagnetic_2008, helmes_mott_2008}.
A bosonic version of the DMFT equations was first derived by scaling the hopping amplitude of condensed and normal bosons differently \cite{byczuk_correlated_2008}.
Later, it was shown \cite{hubener_magnetic_2009,  hu_dynamical_2009, anders_dynamical_2011, proukakis_bosonic_2013} that the same BDMFT equations can be derived via a systematic expansion of the effective action in the reciprocal coordination number $1 / z$ using a uniform scaling of $J \rightarrow J / z$.

One can derive an effective action for the Bose-Hubbard model at each site $j$ by applying the cavity method, which yields the following expression \cite{li_tunable_2011, proukakis_bosonic_2013}:
\begin{eqnarray}
	S^\eff_j &= &\bint \frac{U}{2} \psi_j^* (\tau) \psi_j^* (\tau) \psi_j (\tau) \psi_j (\tau) \nonumber \\
    &- & J \bint \sum_{ \langle j i \rangle } [\psi_j^* (\tau) \langle \psi_i (\tau) \rangle_j  + \langle \psi_i (\tau) \rangle_j^* \psi_j (\tau)] \nonumber \\
	&+ &\bint \bint' \psinambu^\dagger (\tau) \weissG^{-1}_j (\tau, \tau') \psinambu (\tau'),
	\label{eq:realspaceEffectiveAction}
\end{eqnarray}

where $\weissG_j$ denotes the Weiss function.
The expectation value $\langle \dots \rangle_j$ is with respect to the cavity system, which is missing site $j$.
Thus, one has to include a $1 / z$ correction to the condensate order parameters \cite{proukakis_bosonic_2013}, which we implement within RBDMFT using Eq. (\ref{eq:corrcondorder}).
The (double) underscore indicates the use of the Nambu (matrix) notation such that $\psinambu_j = (\psi_j, \psi_j^*)^T$, where the complex fields $\psi_j (\tau)$ ($\psi_j^* (\tau)$) correspond to the respective annihilation $\bop_j$ (creation $\bdag_j$) operator in Eq.\,\eqref{eq:dBHM} in the coherent-state path-integral formulation.

The self-consistency is closed using the lattice Dyson equation \cite{li_tunable_2011}
\begin{equation}
    \label{eq:latticeDyson}
    \mb{\Gnambu}^{-1} (\iw_n) = \mb{\Gnambu}^{-1}_0 (\iw_n) - \mb{\selfnambu} (\iw_n), \\
\end{equation}
where $\mb{\Gnambu}^{-1}_0 (\iw_n)$ is the noninteracting lattice Green's function given by
\begin{equation}
    \label{eq:nonInteractingLatticeDyson}
    \mb{\Gnambu}^{-1}_0 (\iw_n) = (\mu \mb{\idnambu} + \iw_n \sigmanambu_z \otimes \mb{\mathrm{I}}) - \idnambu \otimes \mb{J} - \idnambu \otimes \mb{\epsilon},
\end{equation}
Here $\mb{J}$ is the hopping matrix, the elements $[\mb{J}]_{ij}$ of which are $-J$ if $i, j$ index neighboring sites and zero otherwise.
$\mb{\epsilon}_{ij} = \delta_{ij} \epsilon_j$ is a diagonal matrix containing the disorder potential for each site.
Bold symbols indicate a $N_s \times N_s$ matrix in real space and the combination with the Nambu matrix structure (double underscore) results in a $2 N_s \times 2 N_s$ matrix.
The tensor product ($\otimes$) in Eq.\ \eqref{eq:nonInteractingLatticeDyson} is used to explicitly reflect the matrix structure.

The lattice Green's function is related to the Weiss function in Eq.\ \eqref{eq:realspaceEffectiveAction} at each site $j$ via the local Dyson equation \cite{li_tunable_2011}
\begin{eqnarray}
	\weissG^{-1}_j (\iw_n) = (\Gnambu_{jj})^{-1} (\iw_n) + \selfnambu_j (\iw_n),
    \label{eq:localDyson}
\end{eqnarray}
where $\Gnambu_{jj}$ is a $2 \times 2$ Nambu matrix given by the site diagonal elements of the lattice Green's function.
Within RBDMFT the self-energy is assumed to be local such that $\mb{\selfnambu} (\iw_n)$ is a site diagonal matrix.

We introduce a bosonic version of the Anderson impurity model with parameters chosen such that they reproduce the effective action in Eq.\,\eqref{eq:realspaceEffectiveAction} \cite{proukakis_bosonic_2013}
\begin{eqnarray}
    &\hat{H}^A_j &= - (\mu - \epsilon_j) \hat{n}_j + \frac{U}{2} \hat{n}_j ( \hat{n}_j - 1 ) \nonumber \\ 
	&+&\sum_{l=1}^{N_l} \big ( \bdagnambu_j \Vnambu_{lj} \aopnambu_{lj} + \adagnambu_{lj} \epsnambu_{lj} \aopnambu_{lj} \big)
	- [\mb{J} \tilde{\mb{\phi}}]_j [\bop_j + \bdag_j].
    \label{eq:AndersonHamiltonian}
\end{eqnarray}
Here $\bop_j$ ($\bdag_j$) is the annihilation (creation) operator and $\hat{n}_j = \bdag_j \bop_j$ is the number operator on the $j$th impurity site.
The $\aop_{lj}$ ($\adag_{lj}$) operator acts on the bosonic orbital $l = 1, \dots, N_l$ coupled to the $j$th impurity site.
The elements of the corrected condensate-order parameter vector $[\tilde{\mb{\phi}}]_j$ are inserted in the impurity model as derived in \cite{proukakis_bosonic_2013}.
The adjustable Anderson parameters ($e_{lj}$, $\delta_{lj}$, $V_{lj}$, and $W_{lj}$) are compactly introduced in the $\Vnambu_j$ and $\epsnambu_j$ Nambu matrices defined as
\begin{equation}
    	\epsnambu_{lj} = \mat
	{
		e_{lj} / 2 & \delta_{lj} \\
		\delta_{lj}^* & e_{lj} / 2
	}, \quad
	\Vnambu_l = \mat
	{
		V_{lj} & W_{lj} \\
		W^*_{lj} & V^*_{lj}
	}.\\
\end{equation}
The hybridization function is obtained by integrating out the bath's degrees of freedom in the impurity effective action \cite{proukakis_bosonic_2013}
\begin{eqnarray}
    \hybnambu_j (\iw_n) = \frac{1}{4} \sum_l^{N_l} \Vnambu_{lj}^* (\iw_n \sigmanambu_z - \epsnambu_{lj})^{-1} \Vnambu_{lj},
    \label{eq:AndersonHybridization}
\end{eqnarray}
where $\sigma_z$ denotes the third Pauli matrix.
We are then able to connect the Weiss function in Eq.\ \eqref{eq:localDyson} to the impurity model via 
\begin{equation}
 \weissG_j^{-1} (\iw_n) = \iw_n \sigmanambu_z + (\mu - \epsilon_j) - \hybnambu_j (\iw_n).   
\end{equation}
To include the necessary correction to the condensate order parameter within RBDMFT \cite{proukakis_bosonic_2013} one adjusts the condensate order parameter within the hopping term of the Anderson impurity model as follows:
\begin{equation}
\label{eq:corrcondorder}
     [\mb{J} \mb{\tilde{\phi}}]_j = [\mb{J} \mb{\phi}]_j + \phi_j \cdot [\hybnambu_j (0)]_{00} + \phi^*_j \cdot [\hybnambu_j (0)]_{01},
\end{equation}
where $\mb{J} \mb{\phi}$ is the matrix vector product between the hopping matrix and the site-dependent condensate order parameter arranged as a column vector.

The self-energy is then calculated using the Lehmann representation and exact diagonalization of the Anderson impurity Hamiltonian in Eq.\ \eqref{eq:AndersonHamiltonian}.
For bosons this implies that an upper bound for the occupations of the orbitals and the impurity has to be imposed, which limits the numerical accuracy as well as efficiency of the approach \cite{geisler_infinite_2017}.
On the other hand the Lehmann representation allows for a direct computation of the spectral function in the real frequency domain without the need for numerical analytical continuation.

In conclusion we start with a guess for the impurity self-energies as well as condensate order parameters and solve the lattice Dyson equation in Eq.\,\eqref{eq:latticeDyson}.
This allows us to optimize the Anderson parameters via the local Dyson equation in Eq.\,\eqref{eq:localDyson} such that they reproduce the effective action of the lattice model.
The new Anderson parameters can then be used to solve the Anderson Hamiltonian in Eq.\,\eqref{eq:AndersonHamiltonian} by exact diagonalization, which results in new impurity self-energies (and condensate order parameters). This self-consistency is continued until the self-energies (and condensate order parameters) converge to stationary values.

\bibliography{article}

\end{document}